\documentstyle[amscd,amssymb,verbatim,12pt]{amsart}
\def\dspace{\baselineskip=0.3 in}
\begin{document}
\dspace
\title[A Different Approach tO $f(R)$-Cosmology  ]{A Different Approach tO $f(R)$-Cosmology}

\author{\bf S.K.Srivastava}
{ }
\maketitle
\centerline{ Department of Mathematics,}
 \centerline{ North Eastern Hill University,}
 \centerline{  Shillong-793022, India}
\centerline{ srivastava@@.nehu.ac.in; sushilsrivastava52@@gmail.com}

\vspace{1cm}

\centerline{\bf Abstract}

Here $f(R)$-cosmology is discussed using a different approach. This model explains early-inflation, emergence of cosmic background radiation at the exit from inflation, cosmic deceleration during radiation-dominance followed by deceleration due to curvature-induced matter and acceleration in the very late universe due to curvature-induced phantom dark energy. This model predicts collapse in the future universe. Further, a possible avoidence of collapse as well as revival of very early universe is suggested. 

\vspace{0.5cm}

\centerline {\underline{\bf 1. Introduction}}
 
 Observation of cosmic acceleration in the very late universe \cite{sp, ag} created lot of sensation in the arena of cosmology and emerged as the most fundamental theoretical problem. According to Friedmann equations (giving cosmic dynamics), cosmic acceleration is caused due to negative pressure $ p < - \rho/3 $ of the dominant fluid ($ \rho $ being the energy density).  Many field-theoretic, Chaplygin gas \cite[and references therein]{ejc} and $f(R)$-dark enenrgy models \cite{snj, lds, nsd}($R$ being the scalar curvature and $f(R)$ being some function of $R$) were proposed to explain late acceleration and to probe a proper source of the exotic fluid, which is invisible but having gravitational effects. This fluid is popularly known as {\em dark energy}. Here $f(R)$-cosmology is adressed.
 
In $f(R)$-dark energy models, non-linear terms of curvature is identified as DE term . But, a more appropriate theory should begin without imposing any pre-condition or identifying some terms as DE terms {\em a priori}.

 This article presents a different approach to $f(R)$-cosmology, where no curvature term is recognized as gravitational alternative of DE {\em a priori}\cite{sks, sks08c, sks08d} contrary to the approach of refs.\cite{snj, lds, nsd}. 
 
In refs.\cite{snj,lds, nsd}, gravitational equations are derived from the action having Einstein
-Hilbert term and non-linear curvature terms. Terms in gravitational equations, due to non
-linear curvature terms, are recognized as DE terms. In what follows,  trace of $f(R)-$gravitational
 equations are obtained yielding an equation for scalar curvature $R$. In the homogeneous
 space-time, this equation reduces to the second-order equation for the scale factor $ a(t)$. First integral of this differential equation yields the Friedmann equation giving dynamics
 of the universe. Interestingly, the Friedmann equation, obtained so, contains terms for quintessence-like DE term and dark radiation in the early universe as well as dark matter and phantom-like DE term in the late universe. These terms are induced by curvature without identifying any curvature term as DE {\em a priori}. Here DE terms emerge from Einstein term proportional to $R$ as well as non-linear terms of $R$ spontaneously.
 
  The gravitational action of the model being addressed here contains non-linear terms $ R^2$ and $ R^{(2 + r)} $ (with $r > 0$ being a real number). It is interesting to find that dark matter is induced in the present set-up, if $ r = 3.$

The present model yields an interesting cosmological picture from the early universe to the future universe. Here, investigations begin from the Planck scale (the fundamental scale). 
 It is found that the early universe inflates for a short
 period, driven by curvature-induced quintessence dark energy. During this period, elementary particles are produced and lot of energy is released as radiation at the end of inflation. The emitted radiation thermalizes the universe rapidly up to the temperature $ \sim 10^{18}{\rm GeV}$. As a consequence, produced particles due to the decay of curvature-induced quintessence attain the thermal equillibrium with radiation. There are two sources of radiation (i) emitted radiation at the end of inflation and (ii) curvature-induced radiation  recognized as dark radiation. The emitted radiation, is identified with the cosmic  background radiation with very high initial temperature $ \sim 10^{18}{\rm GeV}$. Due to its dominance,  the universe decelerates after exit from inflation heralding the standard cosmology.

 The produced elementary particles, during inflation, undergo various processes of the standard cosmology such as nucleosynthesis,baryosynthesis and hydrogen-recombination, which are not discussed here. Thus, like radiation, we have two types of pressureless matter (i) 
baryonic matter, which is formed due to nucleosynthesis and baryosynthesis of elementary particles produced  and (ii) curvature-induced non-baryonic matter identified as dark matter. After sufficient expansion of the universe, the matter dominates over radiation causing decelerated expansion as $ \sim t^{2/3} $ in the late universe. 

 It is interesting to note that alongwith curvature-induced phantom terms $\rho^{\rm ph}_{\rm de}[1-{(\rho^{\rm ph}_{\rm de})}/{2 \lambda}]$ ($ \rho^{\rm ph}_{\rm de} $ being the phantom energy density) in the late curvature causes another constant $ \lambda $, which is analogous to negative brane-tension in Randall-Sundrum II theory of brane-gravity \cite{rm}. As a remark, it is nice to mention that these type of terms also appear in Freidmann equation based on the loop quantum gravity \cite{ms}. As brane-theory prescriptions are not used here, the curvature-induced $ \lambda $ appearing in this model is identified as {\em cosmic tension} like the Refs. \cite{sks08c}. Curvature-induced phantom dominates at the red-shift $ z = 0.303 $ causing a cosmic jerk. As a consequence, a transition from deceleration to acceleration takes place in the very late universe. It is found that the phenomenon of cosmic acceleration will continue in future too. The accelerating phase will end when phantom density will be equal to twice of the cosmic tension followed by deceleration due to re-dominance of matter.
  
Here it is shown that phantom density will still grow due to expansion of the universe during re-dominance of matter and expansion will stop at time $ t_m \simeq 3.45\times 10^{15} t_0 $ ($ t_0 $ being the present age of the universe). So, the universe will bounce causing contraction in the universe. As a result, matter energy density will increase rapidly such that energy density and pressure density diverge as well as $ a = 0 $ at $ t_{\rm col} \simeq 3.62\times 10^{15} t_0 $. This result predicts collapse of the universe at time $ t_{\rm col}$.

Further, it is discussed that the result of cosmic collapse is obtained using the classical mechanics. It is argued that near $ t_{\rm col}$ energy density and curvature will  be very high. It is analogous to the state of the early universe. So, production of quantum particles will take place due to rapid change in topology of the space-time in the vicinity of $ t_{\rm col}$. It is found that back-reaction of produced particles will avoid the cosmic collapse and universe will expand exponentially when $ t > t_{\rm col} $. 

Natural units $(k_B = {\hbar} = c = 1)$ (where $k_B, {\hbar}, c$ have their
usual meaning) are used here. GeV is used as a fundamental unit and we have $1 {\rm GeV}^{-1}
= 6.58 \times 10^{-25} sec$ and $1 {\rm GeV} = 1.16 \times {10^{13}}^0 K.$ 

\bigskip

\centerline {\underline{\bf 2.  $f(R)$- gravity and Friedmann equations }}

 \smallskip
 The action is taken as
$$ S = \int {d^4x} \sqrt{-g} \Big[\frac{R}{16 \pi G} + \alpha R^2 + \beta
R^{(2 + r)}  \Big],  \eqno(2.1)$$ 
where $G = M_P^{-2} (M_P = 10^{19} {\rm GeV}$ is the Planck mass), $\alpha$ is
a dimensionless coupling constant, $\beta$ is a constant having dimension
(mass)$^{(-2 r)}$ (as $R$ has mass dimension 2) with $r$ being a positive real
number.

The action (2.1) yields gravitational field equations
$$ \frac{1}{16 \pi G} (R_{\mu\nu} - \frac{1}{2} g_{\mu\nu} R) + \alpha (2
\triangledown_{\mu} \triangledown_{\nu} R - 2 g_{\mu\nu} {\Box} R -
\frac{1}{2} g_{\mu\nu} R^2 + 2 R R_{\mu\nu} ) $$
$$ + \beta (2 + r) ( \triangledown_{\mu} \triangledown_{\nu} R^{(1 + r)} -
g_{\mu\nu} {\Box} 
R^{(1 + r)}) + \frac{1}{2}\beta g_{\mu\nu} R^{(2 + r)} $$
$$- \beta (2 + r) R^{(1 +   r)} R_{\mu\nu}   = 0, \eqno(2.2)$$  
where $\triangledown_{\mu}$ stands for the covariant derivative.

Taking trace of (2.2), it is obtained that
$$ - \frac{R}{16 \pi G} - [6 \alpha + 3 \beta (1 + r)(2 + r)  R^r ] {\Box} R -
3 \beta r (1 + r)(2 + r)R^{(r - 1)} \triangledown^{\mu}R \triangledown_{\mu}R$$
$$+ \beta r R^{(2 + r)}  = 0 \eqno(2.3)$$ 
with
$$ {\Box} = \frac{1}{\sqrt{-g}} \frac{\partial}{\partial x^{\mu}}
\Big(\sqrt{-g} g^{\mu\nu} \frac{\partial}{\partial x^{\nu}} \Big). \eqno(2.4)$$

In (2.3), $[6 \alpha + 3 \beta (1 + r)(2 + r)  R^r ]$
emerges as a coefficient of ${\Box} R$ due to 
presence of terms $\alpha R^2$ and $\beta R^{(2 + r)}$ in the action
(2.1). If $\alpha = 
0$, effect of $R^2$ vanishes and  effect of $R^{(2 + r)}$ is switched
off for $\beta = 0$. So, like \cite{sks} an {\em effective} scalar curvature
${\tilde R}$ is defined as 
$$ \gamma {\tilde R}^r = [6 \alpha + 3 \beta (1 + r)(2 + r)  R^r ] , \eqno(2.5)$$
where  $\gamma$ is a constant having dimension (mass)$^{-2r}$ being
used for dimensional correction.

Connecting (2.3) and(2.5), it is obtained \cite{sks08d}that
$$-\frac{1}{16 \pi G} \frac{1}{\gamma {\tilde R}^{r - 1}}\Big[\frac{6
  \alpha}{\gamma {\tilde R}^{r}} - 1 \Big] + {\Box}{\tilde R} +
  (r - 1) {\tilde R}^{-1} \triangledown^{\mu}{\tilde R}
  \triangledown_{\mu}{\tilde R}$$
$$ - (1 - r) \frac{\gamma {\tilde R}^{r - 1}}{6 \alpha - \gamma {\tilde R}^{r}}\triangledown^{\mu}{\tilde R}\triangledown_{\mu}{\tilde R} +
  r {\tilde R}^{-1} \triangledown^{\mu}{\tilde R} 
  \triangledown_{\mu}{\tilde R}$$
$$+ r)(2 + r)/ \gamma^2]
  {\tilde R}^{2 r - 1} \Big[\frac{\gamma {\tilde R}^{r} - 6\alpha}{3 \beta (1 + r)(2 + r)} \Big]^{(1/r + 2)} = 0 . \eqno(2.6)$$ 

Experimental evidences \cite{ad} support
 spatially homogeneous and
flat model of the universe 
$$dS^2 = dt^2 - a^2(t) [dx^2 + dy^2 + dz^2] \eqno(2.7)$$
with $a(t)$ being the scale factor.

For $a(t)$, being the power-law function of cosmic time, ${\tilde R} \sim a^{-n}$. For
example,  ${\tilde R} \sim a^{-3}$ for
matter-dominated model. So, there is no harm in taking
$$  {\tilde R} = \frac{A}{ a^n} , \eqno(2.8)$$
where $n > 0$ is a real number and $A$ is a constant with mass dimension 2. 

Using (2.6), (2.6) is obtained as
$$ \frac{\ddot a}{a} + \Big[2 - n - n (r - 1) + \frac{n (1 - r) \gamma A^r
  a^{- nr}}{6 \alpha - \gamma A^r a^{- nr}} - n r \Big] \Big(\frac{\dot a}{a} \Big)^2 = \frac{a^{n r}}{16 \pi G \gamma A^r} \Big[\frac{6 \alpha a^{nr}}{\gamma A^r} - 1 \Big]$$
$$ - \frac{ \beta^{-1/3} }{n (\gamma A^r a^{-nr})^2[3r(1 + r)(2 + r)]^{1 + 1/r} }[6 \alpha - \gamma A^r a^{-nr}]^{2 + 1/r} . \eqno(2.9)$$

Approximating (2.9) for small $ a(t) $ in the early universe and integrating, we obtain
$$ \Big(\frac{\dot a}{a} \Big)^2 = \frac{B}{a^{(2 + 2M)}} - \frac{2 \beta^{-1/r}}{n (\gamma A^r)^2 [3r(1 + r)(2 + r)]^{1 + 1/r}a^{(2 + 2M)}}$$
$$\times \int a^{(1 + 2M + 2nr)}[6 \alpha - \gamma A^r a^{-nr}]^{2 + 1/r}  \eqno(2.10)$$
with $B$ being the integration constant and
$$ M = 2 - n - n r. \eqno(2.11)$$

Approximation of (2.10)for large $ a(t)$ in the late universe leads to
$$\frac{\ddot a}{a} + \Big[2 - 2nr \Big] \Big(\frac{\dot a}{a} \Big)^2 = D a^{nr} - E a^{2nr} , \eqno(2.12a)$$
where
$$D = \Big(\frac{6\alpha}{\gamma A^r}\Big) \Big[ \frac{1}{16 \pi G n} - (2 +
1/r)\frac{[3 r (1 + r)(2 + r)]^{-1-1/r}}{n}\Big(\frac{6\alpha}{\gamma
  A^r}\Big) \Big] \eqno(2.12b)$$  
and
$$E = \Big(\frac{6\alpha}{\gamma A^r}\Big)^2 \Big[ \frac{1}{16 \pi G n} -
\frac{[3 r (1 + r)(2 + r)]^{-1-1/r}}{n}\Big(\frac{6\alpha}{\gamma A^r}\Big)
\Big]. \eqno(2.12c)$$ 

(2.12a) is integrated to
$$ \Big(\frac{\dot a}{a} \Big)^2 = \frac{B}{a^{(2 + 2N)}} + \frac{2D}{(2 + 2N +
    nr)} a^{nr} \Big[1 - \frac{E(2 + 2N + nr)}{D(2 + 2N + 2nr)}a^{nr} \Big]
  \eqno(2.13)$$  
with
$$ N = 2 - 2nr .\eqno(2.14)$$ 

Further, it is found that if $M = 1$, the first term on r.h.s.(right hand
side) of (2.10) gives radiation. Moreover, the first term of r.h.s. of
(2.13) has the form of matter density if $N = 1/2$  . So, setting $M = 1$ in (2.10) and $N = 1/2$ in (2.14) to get a viable cosmology, it is
obtained that 
$$ n = \frac{1}{4}\eqno(2.15a)$$
and
$$ r = 3. \eqno(2.15b)$$

\bigskip

\centerline {\underline{\bf 3. Power-law inflation, origin of matter and reheating after inflation  }}  

\noindent {\underline{\bf 3(a). Power-law inflation}} The Friedmann equation (2.10) for the early universe is approximated to
$$ \Big(\frac{\dot a}{a} \Big)^2 \approx \frac{B}{a^4} + \frac{8 \pi G}{3}\rho^{\rm qu}\noindent_{\rm de}\eqno(3.1)$$ 
using (2.11) and $ M = 1 $ for
$$a < \Big(\gamma A^3/6 \alpha \Big)^{4/3} = a_c .\eqno(3.2)$$

Using (3.2), the energy density in (3.1)is obtained as
$$ \rho^{\rm qu}_{\rm de} =  \frac{3}{8 \pi G} \Big[\frac{16
  \beta^{-1/3}}{11 (\gamma A^3)^{-1/3} [180]^{4/3}}  a^{3/2}\Big]\Big[ a^{- 3/4} - a_c^{- 3/4} \Big]^{7/3}  \eqno(3.3)$$
is caused by linear as well as non-linear
terms of curvature. Due to its origin from curvature, $ \rho^{\rm qu}_{\rm de} $ is identified as dark energy density.  

Interestingly, a radiation density  term $B/a^4$ emerges in (3.1a)
spontaneously from gravity. This type of term emerged first in brane-gravity inspired Friedmann
equation. So, analogous to brane-gravity, here also $B/a^4$ is called dark radiation. 

Here investigations start at the Planck scale, where DE density is obtained
around $10^{75} {\rm GeV}^4$. Using $ \rho^{\rm qu}_{\rm de} = 10^{75} {\rm GeV}^4$ at $a = a_P$, (3.3) is re-written as
$$\rho^{\rm qu}_{\rm de} = 10^{75}\Big(\frac{ a}{a_P}\Big)^{3/2} \Big[\frac{ a^{- 3/4} - a_c^{-
  3/4}  }{ a_P^{- 3/4} - a_c^{- 3/4}}
  \Big]^{7/3} \eqno(3.4)$$
  
  (3.4) and the conservation equation 
$$ {\dot \rho}_{\rm de} + 3 \frac{\dot a}{a} ( \rho_{\rm
  de} + p_{\rm de} ) = 0 \eqno(3.5)$$
yield
$$p^{\rm qu}_{\rm de} =  - \frac{3}{2}\rho^{\rm qu}_{\rm de} + \frac{7}{12}10^{75}\Big(\frac{ a}{a_P}\Big)^{3/2} \Big[\frac{ a^{- 3/4} - a_c^{-
  3/4}  }{ a_P^{- 3/4} - a_c^{- 3/4}}
  \Big]^{7/3} . \eqno(3.6)$$
  (3.6) shows violation of SEC for $a_P
\le a(t) < a_c$. It implies dark energy given by (3.2) mimics quintessence.
\bigskip

As $a_P$ is expected to be extremely small, so 
$\rho^{\rm qu}_{\rm de}$ dominates over the radiation in (3.1).  Moreover,
(3.6) shows that $\rho^{\rm qu}_{\rm de} = 0$  at $a = a_c$. So, for $a_P <
a(t) < a_c,$ cosmic dynamics is given by
$$ \Big(\frac{\dot a}{a} \Big)^2 \simeq  \frac{8 \pi \times
  10^{37}}{3} \Big(\frac{ a}{a_P}\Big)^{3/2} \Big[\frac{ a_c^{- 3/4} - a^{-
  3/4} }{ a_c^{- 3/4} - a_P^{- 3/4}}  \Big]^{7/3} \simeq  \frac{8 \pi \times
  10^{37}}{3} \Big(\frac{ a}{a_P}\Big)^{-1/4}. 
\eqno(3.7)$$ 
which integrates to
$$ a(t) = a_P \Big[1 + \frac{M_P}{8 \sqrt{3 \pi}}(t - t_P) \Big]^8
\eqno(3.8)$$ 
giving {\em power-law inflation} during  the time period
$$  t_c - t_P \simeq   7.77 \times 10^{4} t_P , \eqno(3.9)$$
 if
$$ \frac{a_c}{a_P} = 10^{28} \eqno(3.10)$$
being required for sufficient inflation.

\smallskip

\noindent {\underline{\bf 3(b). Origin of matter and reheating after inflation }}

Ref. \cite{sks08d} demonstrates in detail that the curvature-induced quintessence DE with density (3.4) can be realised through the quintessence scalar $\phi_0(t)$ with potential
\begin{eqnarray*}
 V(\phi_0) &=& \frac{1}{2} (\rho_{\rm de} - p_{\rm de}) \\&=& \frac{5}{4} F
 [a^{-3/4} - a_c^{-3/4}]^{4/3} \Big[a^{3/2} ([a^{-3/4} - a_c^{-3/4}) -
 \frac{7}{30}\Big] \\&=& \simeq \frac{5}{4} F a_P^{-1/4}  e^{-[\phi_0 M_P^{-1} \sqrt{2 \pi}
]},
\end{eqnarray*}
\vspace{-1.8cm}
\begin{flushright}
(3.11a)
\end{flushright}
where 
$$ a(t) = a_P e^{[\phi_0 M_P^{-1} \sqrt{32 \pi} ]} .  \eqno(3.11b)$$
and
$$ F =  10^{75} {a_P^{-3/2}\Big[ a_P^{- 3/4} - a_c^{- 3/4} 
  \Big]^{-7/3}}.\eqno(3.11c)$$
  
The curvature-inspired $\phi_0(t)$ is the background field deriving inflation and $\phi(t,x)$ playing the role of inflaton  can be
realized as $ \phi(t,{\bf x}) = \phi_0(t) + \delta \phi(t,{\bf x})$ with $\delta
\phi(t,{\bf x})$ being the quantum fluctuation. Here, perturbations in the metric
components are ignored for simplicity. The curvature-induced inflaton $\phi(t,x)$ decays into bosons and fermions due to fast topological changes during inflation \cite[for details]{sks08d}. 

Moreover,  fluctuations $ \delta\phi_0(t) $ around $ \phi = \phi(t_c)$ given by the equation
$$ {\ddot \delta\phi_0(t)} + 3 \frac{\dot a}{a}{\dot \delta\phi_0(t)} +
V^{\prime\prime}(\phi)_{\phi = \phi_0({\tilde t})} \delta \phi_0(t) = 0.
\eqno(3.12a)$$ 
(3.12) yields the solution   
 $$\delta\phi_0(t) \simeq \sqrt{2/\pi
  b}\eta^{-12} cos (b \eta - \pi/2 + 23 \pi/8) \eqno(3.12b)$$ 
with  
  $$ \eta = \Big[1 + \frac{M_P}{8 \sqrt{3 \pi}}(t - t_P) \Big]. \eqno(3.12c)$$ 
(3.12a) shows the release of energy as radiation at the end of inflation due to  fluctuations with decaying amplitude. Density of the released energy at the exit of the universe from inflation is obtained as
$$ V(0) - V(\phi_c) \simeq 10^{75} {\rm GeV}^4  .\eqno(3.13)$$
The emitted radiation, having energy density (3.13), reheats the universe rapidly up to the temperature
$$ T_c = 4.8 \times 10^{18} {\rm GeV}. \eqno(3.14)$$

As a result, created elementary particle are highly relativistic and have thermal equillibrium with the emitted radiation.

\smallskip

\centerline {\underline{\bf 4. Deceleration  and acceleration in late and future universe}}
\smallskip

\noindent {\underline{\bf 4(a).Deceleration driven by radiation }}
The early universe from inflation at $ t = t_c $, when $a = a_c$ and $ \rho^{\rm qu}_{\rm de} = 0$ as it is given by (3.4). As discussed above, the emitted radiation at this epoch reheats the universe upto very high temperature given by (3.14) and radiation-dominated era of the standard model of cosmology is recovered. So, the emitted radiation is identified as the {\em cosmic background radiation}(CMB).

Thus, we
have two sources of radiation (i) CMB and (ii) curvature-induced dark radiation given by (3.1a). As temperature of CMB, obtained here, is very
high, dark radiation too will have thermal equillibrium with CMB. So, energy
density of created particles, dark radiation and CMB together will have energy
density 
$$ \rho_r = 10^{75} \Big(\frac{a_c}{a} \Big)^4. \eqno(4.1)$$

So, at the end of inflation ($a = a_c$),Friedmann equation (3.1a) 
reduces to
$$ \Big(\frac{\dot a}{a} \Big)^2 \simeq \frac{8 \pi M_P^2}{30}\Big(\frac{a_c}{a} \Big)^4, \eqno(4.2)$$
and it is integrated to
$$ a(t) = a_c [1 + 4 M_P \sqrt{ \pi/15 a_c^4} (t - t_c) ]^{1/2}. \eqno(4.3)$$

\noindent {\underline{\bf 4(b).Matter-dominance and deceleration}}
Like radiation, we
have two types of matter also (i) dark matter given as $\frac{ 3 C}{8 \pi G
  a^3}$ on setting $ N = 1/2 $ in (2.13) 
  , which is non-baryonic due to its origin from gravity and (ii) baryonic matter formed by
elementary particles ( produced during inflation) through various
processes of 
standard cosmology such as nucleosynthesis, baryosynthesis and recombination
of hydrogen not being discussed here.

 It is interesting to see that if
$$ \rho^{\rm ph}_{\rm de} = \frac{D}{5 \pi G} a^{3/4} \eqno(4.4)$$
and
$$\lambda = \frac{3 D^2}{25 \pi GE},  \eqno(4.5)$$
(2.13) looks like
$$ \Big(\frac{\dot a}{a}\Big)^2 =  \frac{0.27 H^2_0}{a^3} + 0.73 H_0^2 a^{3/4}
\Big\{1 - \frac{0.73 \rho_0^{\rm cr} a^{3/4}}{2 \lambda} \Big\} \eqno(4.6)$$
using current values of matter density $\rho^{(m)}_0 = 0.27\rho_0^{\rm cr}$ and dark energy density $\rho^{\rm ph}_{{\rm de}0} = 0.73\rho_0^{\rm cr}$ provided by WMAP \cite{abl}. Here $\rho_0^{\rm cr} = \frac{3 H_0^2}{8 \pi G}$ with $H_0 = 100h km/Mpc second = 2.32
\times 10^{-42} h {\rm GeV} = [0.96 t_0]^{-1}$ 
with $t_0 = 13.7 {\rm Gyr} = 6.6 \times 10^{41} {\rm GeV}^{-1}$ and $h = 0.68$.

The conservation equation (3.5) for $\rho^{\rm ph}_{\rm de}$ yields
$$ {\rm w}^{\rm ph}_{\rm de} = - \frac{5}{4}. \eqno(4.7)$$
It means that the curvature-induced energy density $\rho^{\rm ph}_{\rm de}$
mimics phantom . Thus, in the
late universe, a
phantom model is obtained from curvature without using any other source of exotic
matter alongwith cosmic tension $ \lambda $(mentioned in the introduction). 

From (4.6), we find that ${0.27 H^2_0}/{a^3}> 0.73 H_0^2 a^{3/4}$ for $ a < a_* $ and ${0.27 H^2_0}/{a^3}< 0.73 H_0^2 a^{3/4}$ for $ a > a_* $. It means that a transition takes place at
$$ a = a_* = \Big(\frac{23}{73}\Big)^{4/15}     = 0.767. \eqno(4.7)$$
It shows that phantom terms in (4.6) dominates over matter term  at red-shift
$$ z_* = \frac{1}{a_*} - 1 = 0.303,  \eqno(4.8)$$
which is very closed to lower limit of $z_*$ given by 16 Type supernova observations \cite{ag}.
\vspace{0.5cm}

For $ a < 0.767$, $[{0.73 \rho_0^{\rm cr} a^{3/4}}]^2 << {0.73 \rho_0^{\rm cr} a^{3/4}}$. So, (4.6) is approximated as
$$ \Big(\frac{\dot a}{a}\Big)^2 = \frac{0.27 H^2_0}{a^3}, \eqno(4.9) $$
which integrates to
$$a(t) = a_d [1 + \frac{3}{2} \sqrt{0.27} H_0 a_d^{-3/2}(t - t_d)]^{2/3}
. \eqno(4.10)$$
Here, $t_d = = 386
{\rm kyr} = 2.8 \times 10^{-5} t_0 $ is the decoupling time ( decoupling of matter from radiation) and
the scale factor $a_d$ at $t = t_d$ is given by $1/a_d = 1 + z_d = 1090$ (WMAP results). (4.10) shows deceleration during matter-dominance. 

\smallskip

\noindent {\underline{\bf 4(c).Late acceleration during phantom dominance}} For $ a \ge 0.735$, (4.6) is approximated as
$$\Big(\frac{\dot a}{a} \Big)^2 = 0.73 H_0^2 a^{3/2} \Big[ a^{-3/4} -
  \frac{0.73 \rho^0_{\rm cr} }{2 \lambda} \Big\}\Big]    \eqno(4.11)$$
with $H_0$ given above 

(4.11) integrates to
\begin{eqnarray*}
 a(t) &=&  \Big[ \frac{0.73 \rho^0_{\rm cr}} {2 \lambda} + 
\Big\{\sqrt{ 1.22 - \frac{0.73 \rho^0_{\rm cr}} {2 \lambda}}
\\ &&
- \frac{3}{8} H_0 \sqrt{0.73} (t - t_*)
\Big\}^2 \Big]^{- 4/3}
\end{eqnarray*}
\vspace{-1.7cm}
\begin{flushright} 
(4.12)
\end{flushright}
as $ a_*^{-3/4} = 1.22$. (4.12)shows acceleration and it is singularity-free. 

\smallskip

\centerline {\underline{\bf 5. Re-dominance of matter, cosmic collapse and its avoidance      }}

\noindent {\underline{\bf 5(a). Deceleration in future universe and cosmic collapse}}

(4.6)shows that phantom-dominance will end at $ a(t_e) =a_e = [{0.73 \rho^0_{\rm cr} }/{2 \lambda}]^{-4/3}$ ie. $ \rho^{\rm ph}_{\rm de} = 2 \lambda $and matter will re-dominate. So,(4.6) will again reduce to (4.9) and the future universe will decelerate as     
$$a(t) = a_e [1 + \frac{3}{2} ]\sqrt{0.27} H_0 a_e^{-3/2}(t - t_e)]^{2/3}
 \eqno(5.1)$$ 

It is notable that phantom density  increases as universe expands, so $ \rho^{\rm ph}_{\rm de} > 2 \lambda $ when $ a > a_e $. As a consequence, $ \rho^{\rm ph}_{\rm de} 
\{1 - \rho^{\rm ph}_{\rm de} /{2 \lambda} \} < 0 $ using (4.4)in (4.6). As growth of $\rho^{\rm ph}_{\rm de}$ will continue, at a certain value $a_m > a_e$ of $a(t)$
$$\frac{0.27 H^2_0}{a_m^3} = 0.73 H_0^2 a_m^{3/4}
\Big\{ \frac{0.73 \rho_0^{\rm cr} a_m^{3/4}}{2 \lambda} - 1\Big\} .\eqno(5.2)$$
(5.2) shows that ${\dot a} = 0$ at $ a = a_m $. So, universe will bounce back and contract. As a consequece, dominance of matter will increase and FE will have the form of (4.9) yielding
$$ H = \Big(\frac{\dot a}{a} \Big) \simeq - \sqrt{0.27} H_0 a^{-3/2} . \eqno(5.3a)$$
(5.3a) integrates to 
$$a(t) = a_m [1 - \frac{3}{2} \sqrt{0.27} H_0 a_m^{-3/2}(t - t_m)]^{2/3}
 \eqno(5.3b)$$ 
 showing decelerated contraction  as ${\ddot a} < 0.$

(5.3b) yields $a(t) = 0$ at
$$ t = t_{\rm col} = t_m +  \frac{2}{3\sqrt{0.27}} H_0^{-1} a_m^{3/2} = 3.62\times 10^{15} t_0 .\eqno(5.4)$$

So, at $ t = t_{\rm col}$, dominating energy density term 
$$ \rho^{(mat)} = 0.27 \rho_0^{\rm cr}/ a^3,  \eqno(5.5)$$ 
in (4.9), will be infinite. These results show cosmic collapse at $ t = t_{\rm col}$.

\noindent {\underline{\bf 5(b).Avoidance of cosmic collapse }} Near the collapse time $ t = t_{\rm col}$, energy density will be very high . So, like curvature-induced quintessence scalar in section3, the energy density (5.5) can be realised through another scalar $ \Phi(t,{\bf x}) $. Due to decay of $ \Phi(t,{\bf x} $ in rapidly changing space-time, particles will be created near the collapse time. Energy density of created particles is obtained as \cite[for details]{sks08d}
$$\rho_{\rm created}  = \frac{680}{27} \sqrt{0.27} \pi^3 a_m^{7/2}
 \Big(\frac{H_0}{V} \Big) e^{3 \pi \sqrt{7}} sinh^3(3 \pi \sqrt{7}/2)
 ({\tilde   \eta}_1)^{- 17/3} a^{-5}.\eqno(5.6)$$
 
 It is natural to think that created  particles will effect cosmic
dynamics. As a consequence, 
FE (4.9) is modified as
$$ H^2  \simeq {0.27} H_0^2 a^{-3} + \frac{8 \pi}{3} M_P^{-2} \rho_{\rm created} \eqno(5.7)$$ 

The solution of (5.7) can be taken as
$$ a = a_{\rm col} exp [|D (t_{\rm col} - t)\}| + \gamma |D (t_{\rm
  col} - t)|^2],  \eqno(5.8)$$
where $a_{\rm col} = a(t = t_{\rm col})$, $D$ is a constant of mass dimension and constant $\gamma$ is dimensionless. 

If (5.8) satisfies (5.7),
$$\frac{3}{8 \pi} M_P^2 D^2 = \frac{0.81}{8 \pi} M_P^2 H_0^2 a_{\rm col}^{-3}
+ X a_{\rm   col}^{-5} ,\eqno(5.9a)$$
$$ \frac{3}{2 \pi} M_P^2 D^2 \gamma
 = - 3 \frac{0.81}{8 \pi} M_P^2 H_0^2 a_{\rm col}^{-3} - 5 X a_{\rm   col}^{-5}, \eqno(5.9b)$$
 $$ \frac{3}{2 \pi} M_P^2 D^2
 \gamma^2  
 = \Big[- 3 \gamma + \frac{9}{2} \Big] \frac{0.81}{8 \pi} M_P^2 H_0^2 a_{\rm
   col}^{-3} + \Big[- 5 \gamma + \frac{25}{2} \Big] X a_{\rm   col}^{-5}\eqno(5.9c)$$
 where
 $$ X = \frac{680}{27} \sqrt{0.27} \pi^3 a_m^{7/2}
 \Big(\frac{H_0}{V} \Big) e^{3 \pi \sqrt{7}} sinh^3(3 \pi \sqrt{7}/2)
 ({\tilde   \eta}_1)^{- 17/3}.$$

(5.9a) and (5.9b) yield $ \gamma  = - \frac{15}{32} .$ At the largest energy mass scale i.e. the Planck mass, energy density is obtained as $M_P^4/8 \pi^2$. Using it in (5.9a,b,c),$ a_{\rm col} = 2.25 \times 10^{-42} $
and
$ D = \frac{M_P}{\sqrt{3 \pi}} $  
are obtained after some manipulations.

\smallskip

\centerline {\underline{\bf 6. Salient features and concluding remarks}}

\smallskip
Here, a cosmological picture is obtained from the gravitational action containing the linear Einstein term as well as non-linear terms $ R^2 $ and $ R^5 $ using an approach different from the work \cite{snj,lds, nsd}.  This approach has an advantage to have power to explain(i) power-law inflation in the early universe and graceful exit from this phase,  (ii) creation of SM particles,  (iii) recovery of the standard cosmology with the cosmic background radiation with extremely high initial temperature $ \sim 10^{18}{\rm GeV}$,  (iv) deceleration of the universe driven by emitted radiation during the inflationary phase and particles in thermal equilibrium with radiation,(v) deceleration driven by curvature induced dark matter and baryonic matter caused by various processes like nucleosynthesis, baryosynthesis, hydrogen re-combination of elementary particles created during inflation, (vi) dominance of curvature-induced phantom at red-shift $z = 0.303$ (which is consistent with observational results), (vii) transient acceleration driven by phantom in the very late universe, (viii) re-dominance of matter, contraction of the universe, collapse of the universe at time $t_{\rm col}= 3.62\times 10^{15} t_0 $ , (ix) avoidance of collapse due to creation of particles near the time $t_{\rm col}$ as well as its back-reaction and  (x) rebirth of the universe after $t_{\rm col}$.

Thus, this model predicts the possible revival of the state of the early universe at time $t_{\rm col}= 3.62\times 10^{15} t_0 $ .


\begin{thebibliography}
\smallskip
\bibitem{sp}
 S. J. Perlmutter $et$ $al.$, Astrophys. J. {\bf 517},(1999)565;
 astro-ph/9812133;  D. N. Spergel $et$ $al$,  Astrophys J. Suppl. {\bf 148}
 (2003)175[ astro-ph/0302209] and references therein.
 
 \smallskip
\bibitem{ag}
 A. G. Riess $et$ $al$, Astrophys. J. {\bf 607}, (2004) 665 [ astro-ph/0402512].

\smallskip
\bibitem{ejc}
 E.J.Copeland, M.Sami and S. Tsujikawa, Int. J. Mod. Phys. D, {\bf
 15},(2006)1753 [hep-th/0603057] and references therein. 
 
\smallskip
\bibitem{snj}
S. Nojiri and S.D.Odintsov, Int.J. Geom. Meth. Mod. Phys. {\bf 4},(2007)115
[hep-th/0601213 ]and references therein; O.M.Lecian and G. Montani,arXiv:0807.4428 [gr-qc].

\smallskip
\bibitem{lds}
L. Amendola, D. Polarski and S. Tsujikawa, Phys.Rev. Lett. {\bf 98} (2007)
131302 [astro-ph/0603703] ; L. Amendola, D. Polarski, R.Gannouji and S. Tsujikawa,  Phys.Rev.D,
{\bf 75} (2007) 083504 [gr-qc/0612180]; S.Capozziello et al, astro-ph 0604431 ;S. Nojiri and S D Odintsov,
hep-th 0608008 .

\smallskip
\bibitem{nsd}
S. Nojiri and S.D.Odintsov,arXiv:0801.4843[astro-ph]; arXiv:0807.0685[hep-th].

\smallskip
\bibitem{sks}
S.K.Srivastava, astro-ph/0511167;astro-ph/0602116; Int.J.Mod.Phys.A {\bf 22
  (6)} (2007)1123 [hep-th/0605019]; 
 Phys.Lett. {\bf B 643} (2006) 1[astro-ph/0608241];  Phys.Lett. {\bf B 648} (2007) 119[astro-ph/0603601]; 
 Int. J. Mod. Phys. D, {\bf 17(5)}(2008)755 [ astro-ph/0602116]; 
Int. J. Theo. Phys. {\bf 47} (2008) 1986
[arXiv:0706.0410 [hep-th]].

 \smallskip
\bibitem{sks08c}
S.K.Srivastava, arXiv:0802.0967[gr-qc].

\smallskip
\bibitem{sks08d}
S.K.Srivastava, arXiv:0809.1950 [gr-qc].
  
\smallskip
\bibitem{rm}
 R. Maartens, Living Rev. Rel. {\bf 7} (2004) 7 [gr-qc/0312059].
 
 \smallskip
\bibitem{ms}
M. Sami, P. Singh and S. Tsujikawa, Phys. Rev.D, {\bf 74} (2006)043514[gr-qc/0605113].

\smallskip
\bibitem{ad}
A.D. Miller $et$ $al$ , Astrophys. J. Lett. {\bf 524} (1999) L1; P. de
Bernadis $et$ $al$ , Nature (London){\bf 400} (2000) 955; A.E. Lange $et$ $al$
, Phys. Rev.D{\bf 63} (2001) 042001; A. Melchiorri $et$ $al$ ,
Astrophys. J. Lett. {\bf 536} (2000) L63; S. Hanay $et$ $al$ , 
Astrophys. J. Lett. {\bf 545} (2000) L5.

\smallskip
\bibitem{abl} 
A.B. Lahnas, N.E. Mavromatos and D.V. Nanopoulos,  Int. J. Mod. Phys. D, {\bf 
  12(9)}, 1529 (2003). 
\end{thebibliography}
\end{document}